\documentclass[prd,preprint,aps,dvips,showpacs]{revtex4}
\usepackage{graphicx}
\usepackage[english]{babel}
\usepackage{amscd}
\usepackage{epsfig}
\usepackage{tabularx}
\usepackage{latexsym}
\usepackage{amsmath}
\usepackage{amsfonts}
\usepackage{amssymb}
\usepackage[latin1]{inputenc}
\usepackage{epsfig}
\usepackage{times}
\usepackage[T1]{fontenc}
\usepackage{graphics}
\usepackage{verbatim}
\usepackage[absolute]{textpos}
\usepackage{wrapfig,times}
\usepackage{amsthm}
\usepackage{setspace}

\newcommand{\be}{\begin{equation}}
\newcommand{\ee}{ \end{equation}}
\newcommand{\ben}{\begin{eqnarray}}
\newcommand{\een}{\end{eqnarray}}

\begin{document}

\title{Hermite polynomials and Fibonacci Oscillators}

\author{Andre A. Marinho$^{1}$, Francisco A. Brito$^{1,2}$}

\affiliation{$^{1}$ Departamento de Física, Universidade Federal de
Campina Grande, 58109-970 Campina Grande, Paraiba, Brazil
\\
$^{2}$ Departamento de Física, Universidade Federal da Paraíba,
Caixa Postal 5008, 58051-970 João Pessoa, Paraíba, Brazil}
\date{\today}

\begin{abstract} 
We compute the ($q_1, q_2$)-deformed Hermite polynomials by replacing the quantum harmonic 
oscillator problem to Fibonacci oscillators. We do this by applying the $(q_1,q_2)$-extension of Jackson derivative. The 
deformed energy spectrum is also found in terms of these parameters. We conclude that the 
deformation is more effective in higher excited states. We conjecture that this achievement may 
find applications in the inclusion of disorder and impurity in quantum systems. The ordinary 
quantum mechanics is easily recovered as $q_1 = 1$ and $q_2\to 1$ or vice versa.
\end{abstract}


\maketitle


\section{Introduction}

It is well known that the resolution of the Schrödinger equation leads to the knowledge of 
the temporal and spatial evolution of the form of the wave associated with a non-relativistic 
particle \cite{sak,gri}. This is the Schrödinger picture for the non-relativistic quantum mechanics. 
Several methods and techniques have been developed over the past decades for analytical 
approximate and exact solutions to improve the understanding of its dynamical behavior.

The insertion of $q$-algebra \cite{lav,hin,ofd,arp,okt,mmc} in quantum mechanics as well as the study of $q$-deformed 
harmonic oscillator have been intensively investigated in the literaure \cite{crl,alo,mse,chai,lav1,ng,beh}. F. H. Jackson 
in his pioneering works introduced the $q$-deformed algebra \cite{jac}, where several aspects of 
investigations played an important role for the understanding and development of such 
an algebra. One of its main ingredients is the presence of a deformation parameter $q$, 
introduced in the commutation relations that define the Lie algebra of the system with the
condition that the original algebra is recovered in the limit {of} $q\to 1$. {The study of} $q$-oscillators by using {the} so-called 
Jackson derivative (JD) {has} been considered in order to determine a generalized 
deformed dynamic in a $q$-commutative phase space \cite{flo}. For this purpose one makes use of 
creation and annihilation operators of $q$-deformed quantum mechanics. 

A new proposal for the $q$-calculation is the inclusion of two distinct deformation parameters 
in some physical applications. Starting with the generalization of $q$-algebra \cite{jac}, 
in \cite{arik} was generalized the Fibonacci sequence. Here, the numbers are in that sequence 
of generalized Fibonacci oscillators, where new parameters ($q_1,q_2$ or $p$, $q$) are introduced.
{One should mention that there are several similar studies in the literature \cite{chak,chu,ssm,bur,aba,buk,amg1,bri} with multi-parameters deformed oscillators that do not necessarily obey the  
Fibonacci properties in {the} sense of the seminal paper by Arik \cite{arik} where the spectrum is given by a generalized Fibonacci sequence.} 
They provide a unification of quantum oscillators 
with quantum groups \cite{bie1,mac,fuc,erz,bor}, keeping the 
degeneracy property of the spectrum invariant under the symmetries of the quantum group. 
The quantum algebra with two deformation parameters may have a greater flexibility when 
it comes to applications in realistic phenomenological physical models \cite{dao,gong}.

In this paper we compute the ($q_1,q_2$)-deformed Hermite polynomials by replacing the 
quantum harmonic oscillator problem to Fibonacci oscillators by changing the ordinary 
derivative to Jackson derivative. The deformed energy spectrum is also found in terms of 
these parameters. The ordinary quantum mechanics is easily recovered as {$q_1 = 1$ and $q_2\to 1$ or vice versa.}

The paper is organized as follows. In Section \ref{alg} we present the $q$-deformed algebra. In 
Section \ref{hpd} we obtain the Hermite polynomials with Fibonacci oscillators and finally, in
Section \ref{con} we make our final comments.

\section{Fibonacci oscillators algebra}
\label{alg}

The generalization of integers usually is given by a sequence. The two well-known ways to describe a 
sequence are the arithmetic and geometric progressions. However, the Fibonacci sequence encompasses 
both. By generalizing this sequence, we get the Fibonacci oscillators, so the spectrum can now be given by 
the Fibonacci integers. The algebraic symmetry of the quantum oscillator is defined by the Heisenberg algebra in terms 
of the annihilation and creation operators $c$, $c^{\dagger}$ respectively, and the number operator $N$, as follows \cite{lav}

\be c_i c_{i}^\dagger - q_1^{2}c_{i}^\dagger c_i = q_2^{2N_i}\qquad\mbox{and}\qquad c_i c_{i}^\dagger - 
q_2^{2}c_{i}^\dagger c_i = q_1^{2N_i},\ee
\begin{equation}[N,c^{\dagger}] = c^{\dagger}, \qquad\qquad [N,c] = -c. \end {equation}
In addition, the operators obey the relations 
\begin{eqnarray}c^\dagger c=[N],\;\;\qquad cc^{\dagger} = [1+N], \end {eqnarray}
\be [1+N_{i,q_1,q_2}] = q_1^{2}[N_{i,q_1,q_2}]+q_2^{2N_i},\;\quad\mbox{or}\quad\; [1+N_{i,q_1,q_2}] = q_2^{2}[N_{i,q_1,q_2}]
+q_1^{2N_i}.\ee
The Fibonacci \textit{basic number} is defined as \cite{arik} 
\be \label{eq1}[n_{i,q_1,q_2}] = c_{i}^\dagger c_{i} = \frac{q_1^{2n_i}-q_2^{2n_i}}{q_1^2-q_2^{2}},\ee
where $q_1$ and $q_2$ are parameters of deformation that are real, positive and independent. 
A few $(q_1,q_2)$-numbers are given here:

\begin{equation} [0] = 0, \nonumber\\\end{equation}
\begin{equation} [1] = 1, \nonumber\\\end{equation}
\begin{equation} [2] = q_1^{2}+q_2^{2}, \nonumber\\\end{equation}
\begin{equation} [3] = q_1^4+q_2^{4}+q_1^2q_2^2, \end{equation}
\begin{equation} [4] = q_1^6+q_2^{6}+q_1^2q_2^4+q_1^{4}q_2^{2}, \nonumber\\\end{equation}
\begin{equation} [5] = q_1^8+q_2^{8}+q_1^6q_2^2+q_1^{2}q_2^{6}+q_1^4q_2^4, \nonumber\\\end{equation}
\begin{equation} [6] = q_1^{10}+q_2^{10}+q_1^8q_2^2+q_1^{2}q_2^{8}+q_1^6q_2^4+q_1^4q_2^6, \nonumber\\\end{equation}

\begin{equation} \nonumber{} \cdots\cdots\cdots\cdots\cdots \end{equation}
\begin{equation} [n]! = [1][2][3]\cdots[n].\end{equation}

One may transform the $q$-Fock space into the configuration space (the Bargmann holomorphic representation) \cite{flo} as in the 
following:
\begin{eqnarray} \label{e32}c^{\dagger} = x,\qquad\qquad c = D_{x}^{(q_1,q_2)},\end{eqnarray}
where $D_{x}^{(q_1,q_2)}$ is the Jackson derivative (JD) \cite{jac,bri} defined as
\ben D_{x}^{(q_1,q_2)}f(x)=\frac{q_1^{2}-q_2^{2}}{\ln\left(\frac{q_1^2}{q_2^{2}}\right)}\partial_{x}^{(q_1,q_2)},\qquad\mbox{where}\qquad
\partial_{x}^{(q_1,q_2)}f{(x)}=\frac {f{(q_1^2x)}-f{(q_2^{2}x)}}{x{(q_1^2-q_2^{2})}},\een
such that 
\ben D_{x}^{(q_1,q_2)}f(x)=\frac{f(q_1^2x)-f(q_2^{2}x)}{x\ln\left(\frac{q_1^2}{q_2^2}\right)},\een
and
\ben D_{x}^{(q_1,q_2)^{2}}f(x)=\frac{q_1^2f(q_2^4x)+q_2^2f(q_1^4x)-(q_1^2 + q_2^2)f\left({q_1^2q_2^2x}\right)
}{q_1^4q_2^4x^2\ln\left(\frac{q_1^2}{q_2^{2}}\right)^2},\een
where $ D_{x}^{(q_1,q_2)^{2}} \equiv D_{x}^{(q_1,q_2)}(D_{x}^{(q_1,q_2)})$ \cite{jac}, and so on. {It is worth noting that the 
paper \cite{chak} written by Chakrabarti and Jagannathan was one of the first to show a version of the JD for two parameters, i.e., 
the $(q,p)$-derivative.} 

Using the definition of the $q$-derivative one can easily find several properties of the JD \cite{ext,gas,erns,bon2,kac}, e.g.,  
\be D_x^q(\exp_q(ax))=a\exp_q(ax), \ee
and
\be D_x^q(ax^n)=a[n]x^{n-1},\ee
which will be useful in the calculations that we are going into details shortly. 

{In the present study we work with a deformed algebra governed by two independent parameters. It is interesting to mention that  
we can reduce the pair $(q_1, q_2)$ to just $q$, in order to make a comparison with other studies in the literature with one parameter.  One way of making such a reduction is considering $q_1 = q$ and $q_2 = q^{-1}$. As such, we reduce our \textit{basic number} (\ref{eq1}) to
\be \label{eqt1} [n] = c_{i}^\dagger c_{i} = \frac{q^{2n_i}-q^{-2n_i}}{q^2-q^{-2}},\ee
and consequently we have $q$-oscillators \cite{bur1,dam}.}

\section{Deformed Hermite polynomials}
\label{hpd}

We start with the Schrödinger equation for the harmonic oscillator and introduce Fibonacci oscillators by replacing
the ordinary derivative to Jackson derivative, i.e.,
\ben \label{eq2}-\frac{\hbar^2}{2m}\frac{d^2\Psi}{dx^2}+\frac{m\omega^2x^2}{2}\Psi=E\Psi, \qquad -\frac{\hbar^2}
{2m}D_{x}^{(q_1,q_2)^{2}}\Psi+\frac{m\omega^2x^2}{2}\Psi=E\Psi. \een
We solve this quantum mechanical problem by using the standard power series method 
(analytical method) found in the literature \cite{sak,gri}, 
and for this let us first introduce the dimensionless variable $\xi$
\be \xi=\alpha x=\sqrt{\frac{m\omega}{\hbar}}x.\ee

Now we can write the ordinary Schrödinger equation (\ref{eq2}) in terms of $\xi$ as in the form
\be \label{eq2.1} \frac{d^2\Psi}{d\xi^2} = (\xi^2-K)\Psi,\qquad\mbox{where}\qquad K\equiv\frac{2E}{\hbar\omega}.\ee
In the asymptotic limit ($\xi$ very large) the term $\xi^2$ dominates over the constant term $K$, i.e., 
\be \frac{d^2\Psi}{d\xi^2} \approx \xi^2\Psi,\ee
which has approximate solution, 
\be \Psi(\xi) \approx A\exp\left(-\frac{\xi^2}{2}\right)+B\exp\left(\frac{\xi^2}{2}\right).\ee

This suggest the following Ansatz for the general solution
\be \label{eq2.9} \Psi(\xi) = h(\xi)\exp\left(-\frac{\xi^2}{2}\right).\ee
Now we can get the first and second Jackson derivatives  as follows
\ben \frac{d\Psi}{d\xi} \to D_{\xi}^{(q_1,q_2)} \Psi=\left(D_{\xi}^{(q_1,q_2)} h-\frac{h\xi[2]}{2}\right)
\exp\left(-\frac{\xi^2}{2}\right),\een
\ben \frac{d^2\Psi}{d\xi^2} \to D_{\xi}^{(q_1,q_2)^{2}} \Psi=\left[D_{\xi}^{(q_1,q_2)^{2}} h - [2]\xi D_{\xi}^{(q_1,q_2)} h + 
\left(\xi^2-\frac{[2]}{2}\right)h\right]\exp\left(-\frac{\xi^2}{2}\right).\een
Replacing this into Eq.(\ref{eq2.1}), we obtain 
\be \label{eq2.2} D_{\xi}^{(q_1,q_2)^{2}} h - [2]\xi D_{\xi}^{(q_1,q_2)} h + \left(K-\frac{[2]}{2}\right)h=0.\ee

Many special functions are known as solution to differential equations of the type given in (\ref{eq2.2}). In our particular 
case, the solution {is known in terms} of Hermite polynomials in $\xi$. 
Let us now go into details by proposing a solution in the form of power series in $\xi$, 
\be h(\xi) = a_0+a_1\xi+a_2\xi^2+\cdots = \displaystyle\sum_{j=0}^{\infty}a_j\xi^j.\ee

By applying the first and second JD (or the `$(q_1,q_2)$-derivative') to the series we find respectively
\be D_{\xi}^{(q_1,q_2)}h=a_1+[2]a_2\xi+[3]a_3\xi^2+\cdots = \displaystyle\sum_{j=0}^{\infty}[j]a_j\xi^{j-1},\ee
and
\be D_{\xi}^{(q_1,q_2)^{2}}h=[2]a_2+[2][3]a_3\xi+[3][4]a_4\xi^2+\cdots = \displaystyle\sum_{j=0}^{\infty}([j]+1)([j]+2)a_{j+2}\xi^{j}.\ee
We can now rewrite the Eq.(\ref{eq2.2}), as follows 
\be \displaystyle\sum_{j=0}^{\infty}\left\{([j]+1)([j]+2)a_{j+2}-[2][j]a_j+\left(K-\frac{[2]}{2}\right)a_j\right\}\xi^j=0.\ee
Since the coefficient of each power in $\xi$ should disappear, then
\be ([j]+1)([j]+2)a_{j+2}-[2][j]a_j+\left(K-\frac{[2]}{2}\right)a_j=0,\ee
that is
\be \label{eq2.3} a_{j+2} = \frac{\left([2][j]-K+\frac{[2]}{2}\right)}
{\left([j]+1\right)\left([j]+2\right)}a_j.\ee

For the sake of comparison with the ordinary case, we show that the recursion formula 
(\ref{eq2.3}) gives explicitly the first three coefficients

\be a_2=\frac{\left([2][0]-K+\frac{[2]}{2}\right)}{\left([0]+1\right)\left([0]+2\right)}a_0=
\frac{\left(\frac{[2]}{2}-K\right)}{2}a_0.\ee

\be a_3=\frac{\left([2][1]-K+\frac{[2]}{2}\right)}{\left([1]+1\right)\left([1]+2\right)}a_1=
\frac{\left(\frac{3[2]}{2}-K\right)}{6}a_1.\ee

\be a_4=\frac{\left([2][2]-K+\frac{[2]}{2}\right)}{\left([2]+1\right)\left([2]+2\right)}a_2=
\frac{\left(\frac{2[2]^2+[2]}{2}-K\right)}{[2]^2+3[2]+2}a_2.\ee

The complete solution is written as follows:
\be \label{eq2.95} h(\xi)=h_{even}(\xi)+h_{odd}(\xi)=a_0+a_1\xi+a_2\xi^2+\cdots.\ee
Thus, Eq.(\ref{eq2.3}) determines $h(\xi)$ in terms of the arbitrary constants $a_0$ and $a_1$, a fact that is 
expected for a second-order differential equation. However, some obtained solutions are not 
normalizable. Let us discuss this in details in the following. 

For very large $j$, the recursion formula will be approximately by, 
\be a_{j+2}\approx \frac{[2]}{[j]} a_j,\qquad a_j\approx \frac{C}{([j]/[2])!},\ee
where $C$ is a constant, and with large values of $\xi$, we have
\ben \label{eq2.91} h(\xi)\approx C\sum{\frac{[1]}{([j]/[2])!}\xi^j}\,\approx \,C\sum{\frac{[1]}{[j]!}\xi^{2j}}\,\approx\, 
C\exp_{q_1,q_2}(\xi^2),\een
where $q$-exponential function is defined as \cite{bon2,kac}
\be \exp_q(ax)=\displaystyle\sum_{n=0}^{\infty}{\frac{a^n}{[n]!}x^n},\qquad\mbox{or}\qquad \exp_{q_1,q_2}(ax)=
\displaystyle\sum_{n=0}^{\infty}{\frac{a^n}{[n_{q_1,q_2}]!}x^n}.\ee

Returning to Eq.(\ref{eq2.9}), where we have the asymptotic behavior, and using Eq.(\ref{eq2.91}) where we have that 
$h$ behaves as $\exp_{q_1,q_2}(\xi^2)$, so $\Psi$ behaves as $\exp(\xi^2/2)$ for instance when {$q_1 = 1$ and $q_2\to 1$ 
(or vice versa)}, which is precisely the solution we disregarded since the very beginning. These are types of non-normalizable solutions.

In order to obtain normalizable solutions, the power series  must terminate. This must 
happens in the highest $j$ that we call $n$, so that Eq.(\ref{eq2.3}) produces
\be a_{n+2}=0, \qquad\mbox{$a_1=0$ for $n$ even and $a_0=0$ for $n$ odd}.\ee
Physically acceptable solutions require that Eq.(\ref{eq2.3}) gives
\ben K=[2][n_{q_1,q_2}]+\frac{[2]}{2}\qquad\mbox{or}\qquad K = [n_{q_1,q_2}]+[n_{q_1,q_2}+1],\een
for some non-negative integer $n_{q_1,q_2}$, i.e., the energy must be
\be \label{eq2.00} E_{n_{q_1,q_2}} = \frac{\hbar\omega}{2}\Big([n_{q_1,q_2}]+[n_{q_1,q_2}+1]\Big),
\ee
and when {$q_1=1$ and $q_2\to1$ (or vice versa)}, we have 
\be \label{eq2.01}E_n = \frac{\hbar\omega}{2}\Big(2n+1\Big).\ee

We now focus on determining the ($q_1, q_2$)-deformed Hermite polynomials. The Hermite 
polynomials are a sequence of orthogonal polynomials that arise in probability theory and 
physics \cite{tsc}. As we know, they give rise to the eigenstates of the ordinary (undeformed) 
quantum harmonic oscillator.

For the allowed values of $K$, we have the formula of recursion: 
\be \label{eq2.35} a_{j+2} = \frac{[2]\left([j]-[n]\right)}
{\left([j]+1\right)\left([j]+2\right)}a_j.\ee

Let us now obtain the first three terms of the series (\ref{eq2.95}). If $n = 0$ we have only one term, 
and we must choose $a_1 = 0$ to neutralize $h_{odd}$, and for $j = 0$ in Eq.(\ref{eq2.35}), we obtain $a_2 = 0$,
such that

\be h_0(\xi)=a_0,\qquad\mbox{and}\qquad \Psi_0(\xi)=a_0\exp\left(-\frac{\xi^2}{2}\right).\ee

For $n=1$, $a_0=0$ and $j=1$, one finds $a_3=0$, then 
\be h_1(\xi)=a_1\xi,\qquad\mbox{and}\qquad \Psi_1(\xi)=a_1\xi\exp\left(-\frac{\xi^2}{2}\right).\ee

For $n=2$ and $j=0$, we have $a_2=-\frac{[2]^2}{2}a_0$, and $j=2$ yields $a_4=0$, thus
\be h_2(\xi)=a_0\left(1-\frac{[2]^2}{2}\xi^2\right),\qquad\mbox{and}\qquad \Psi_2(\xi)=a_0\left(1-\frac{[2]^2}{2}\xi^2\right)
\exp\left(-\frac{\xi^2}{2}\right).\ee
In general, $h_n(\xi)$ will be a polynomial of degree $n$ in $\xi$, for $n$ being either even or odd. 
Up to the general factor ($a_0$ or $a_1$) we will call them ($q_1, q_2$)-Hermite polynomials, 
$H_{n}^{(q_1, q_2)}(\xi)$. Since the Eq.(\ref{eq2.2}) is homogeneous, the Hermite polynomials are defined up to 
a multiplicative constant. Adopting the same usual convention of the ordinary (undeformed) case, we choose 
the constants $a_0$ or $a_1$ so that the coefficient of the highest term $\xi^n$ in $h_n(\xi)$ is $[2]^n$. 
This completely defines the other coefficients from the recursion relation (\ref{eq2.35}) by using the allowed 
values of $K$.

We can now write the $(q_1,q_2)$-deformed stationary states for the Fibonacci oscillators as
follows
\be \Psi_{n}^{(q_1,q_2)}(\xi)=A_{n}^{(q_1,q_2)}H_{n}^{(q_1, q_2)}(\xi)\exp\left(-\frac{\xi^2}{2}\right).\ee
By considering the orthogonality relations
\ben \label{eq2.03}\displaystyle\int_{-\infty}^{\infty}{H_{n}^{2(q_1, q_2)}(\xi)\exp(-\xi^2)}d\xi=\sqrt{\pi}2^n
[n_{q_1,q_2}]!,\een
and
\ben \label{eq2.04}\displaystyle\int_{-\infty}^{\infty}{H_{n}^{(q_1, q_2)}(\xi)H_{m}^{(q_1, q_2)}(\xi)\exp(-\xi^2)}
d\xi=0,\qquad\mbox{$n\neq m$},\een
we can determine the constant $A_{n}^{(q_1,q_2)}$ by normalizing $\Psi_{n}^{(q_1,q_2)}(\xi)$, that is
\be \label{eq2.02}\displaystyle\int_{-\infty}^{\infty}{|\Psi_{n}^{(q_1,q_2)}(\xi)|^2}d\xi={|A_{n}^{(q_1,q_2)}|^2}
\displaystyle\int_{-\infty}^{\infty}{H_{n}^{2(q_1, q_2)}(\xi)\exp(-\xi^2)}d\xi=1.\ee

Finally, we have that the $(q_1,q_2)$-deformed stationary states for the Fibonacci 
oscillators are
\be\label{eq2.4} \Psi_{n}^{(q_1,q_2)}(\xi)=
\frac{1}{\sqrt{2^n[n_{q_1,q_2}]!}}
H_{n}^{(q_1,q_2)}(\xi)\exp\left(-\frac{\xi^2}{2}\right),\ee
where the first $(q_1,q_2)$-Hermite polynomials are given by 
\be H_0=1,\nonumber\\\ee
\be H_1=[2]\xi ,\nonumber\\\ee
\be H_2=([2]\xi)^2-2,\nonumber\\\ee
\be \label {eq2.0} H_3=([2]\xi)^3-3[2]^2\xi,\ee
\be H_4=3[2]^2-([2]\xi)^4-12([2]\xi)^2,\nonumber\\\ee
\be H_5=15[2]^3-10[2]^4\xi^3+([2]\xi)^5.\nonumber\\\ee
We can also write these polynomials through the $(q_1,q_2)$-deformed version of the well-known 
Rodrigues formula \cite{sak,gri},
\be \label{eq2.49} H_{n}^{(q_1,q_2)}(\xi)=(-1)^n\exp_{q_1,q_2}(\xi^2)D_{\xi}^{n(q_1,q_2)}\exp_{q_1,q_2}(-\xi^2).\ee

\begin{figure}[htb!]
\centerline{
\includegraphics[{angle=90,height=7.0cm,angle=270,width=7.0cm}]{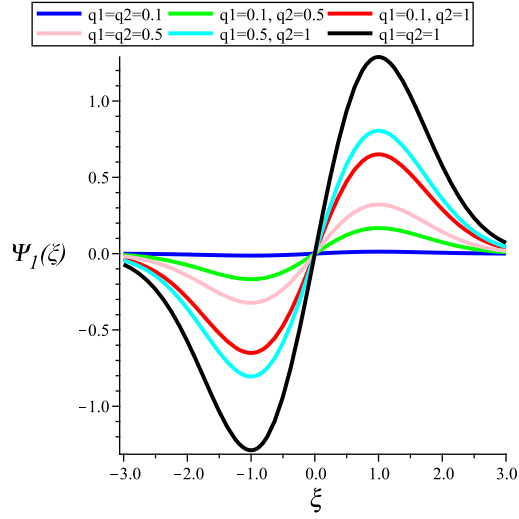}}
\caption{\small{Stationary state $\Psi_1$ for several $(q_1,q_2)$ parameters. The undeformed case is $q_1=1$ and $q_2\to 1$ 
(black curve) and the most deformed case is $q_1=q_2=0.1$ (blue curve).}}
\label{g1}
\end{figure}
\begin{figure}[htb]
\centerline{
\includegraphics[{angle=90,height=7.0cm,angle=270,width=7.0cm}]{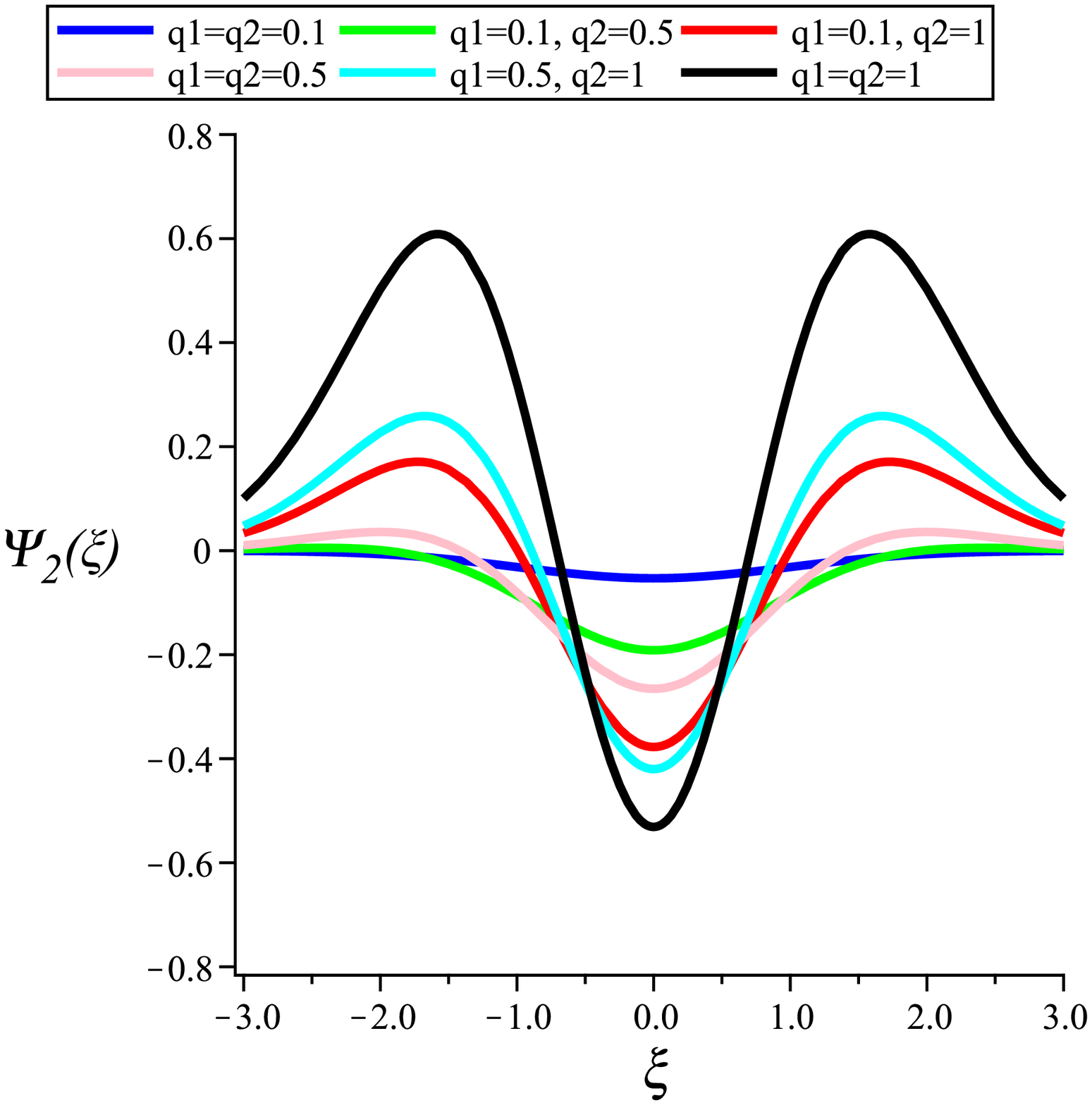}
\includegraphics[{angle=90,height=7.0cm,angle=270,width=7.0cm}]{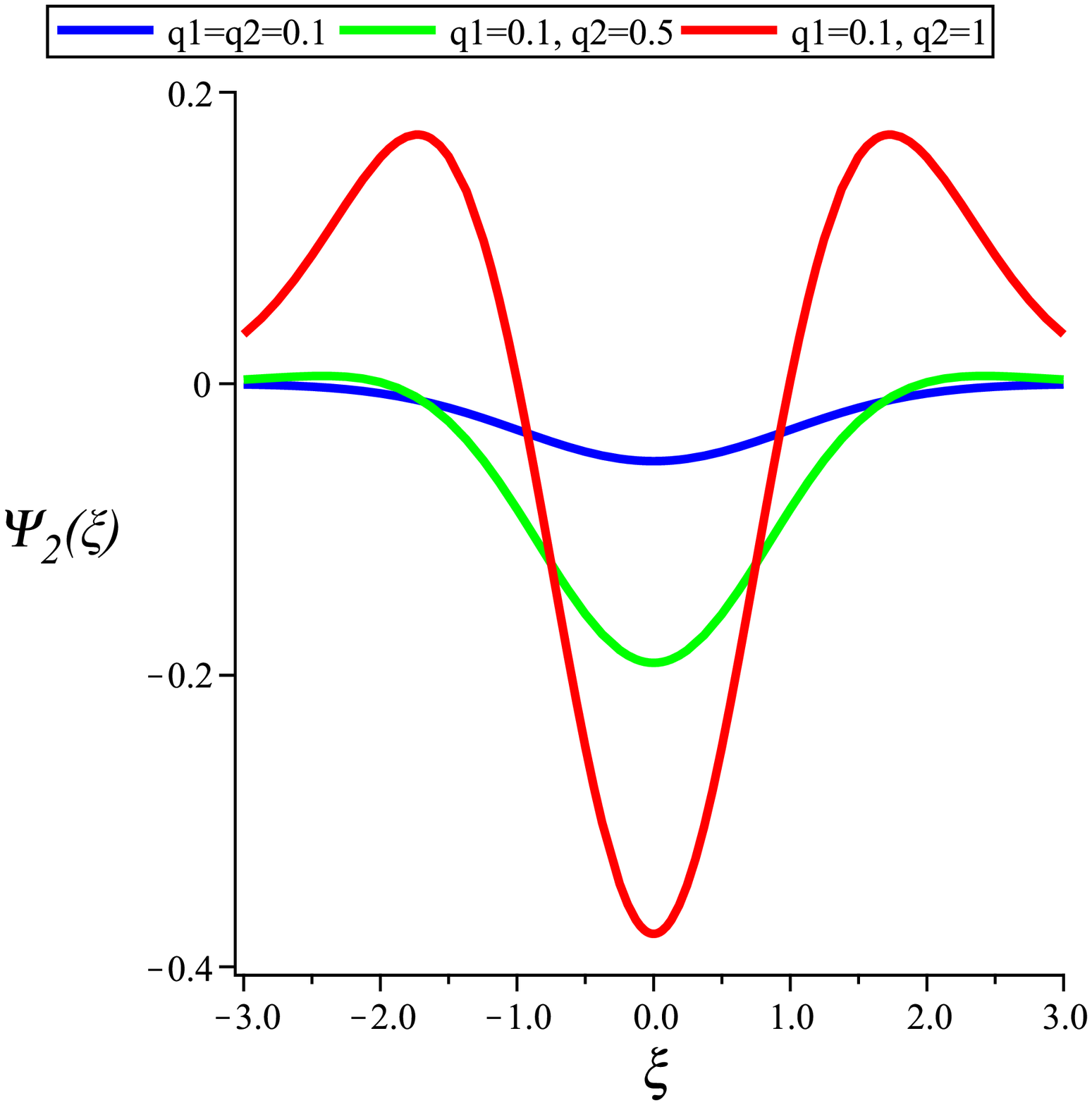}
}\caption{\small{Stationary state $\Psi_2$ for several $(q_1,q_2)$ parameters. The undeformed case is $q_1=1$ and $q_2\to 1$ 
(black curve) and the most deformed case is $q_1=q_2=0.1$ (blue curve).}}
\label{g2}
\end{figure}
\begin{figure}[htb!]
\centerline{
\includegraphics[{angle=90,height=7.0cm,angle=270,width=7.0cm}]{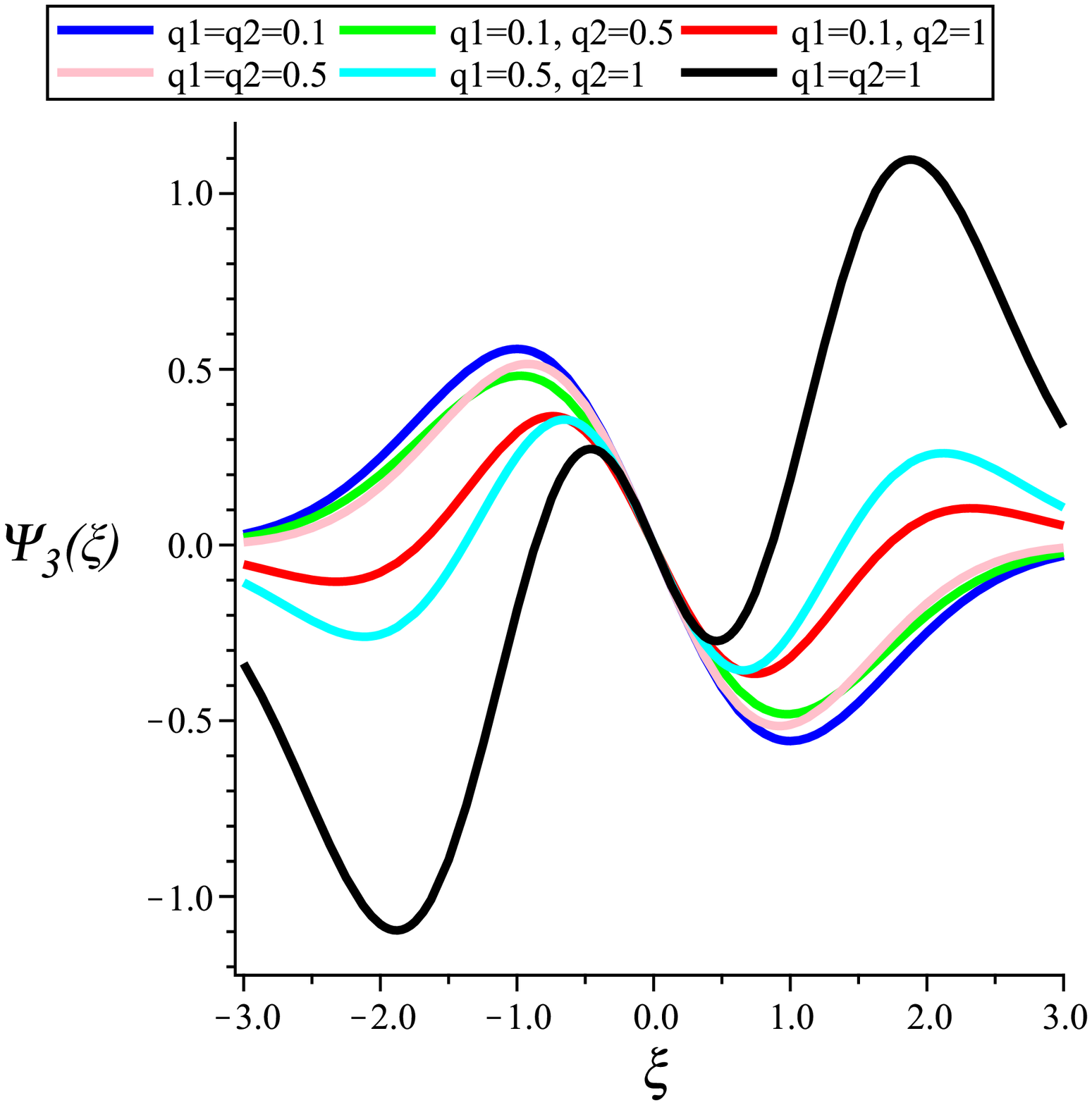}
\includegraphics[{angle=90,height=7.0cm,angle=270,width=7.0cm}]{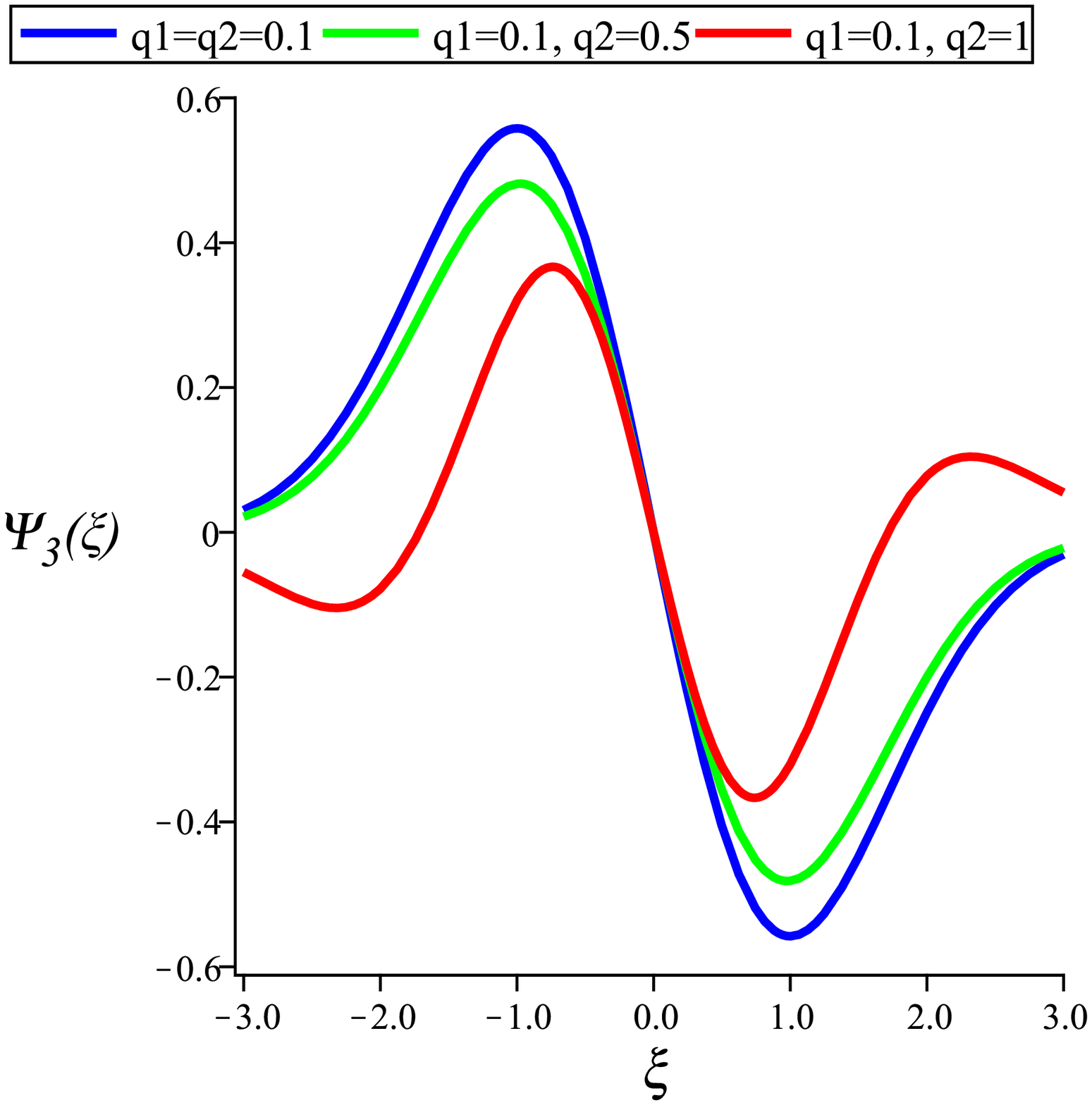}
}\caption{\small{Stationary state $\Psi_3$ for several $(q_1,q_2)$ parameters. The undeformed case is $q_1=1$ and $q_2\to 1$ 
(black curve) and the most deformed case is $q_1=q_2=0.1$ (blue curve).}}
\label{g3}
\end{figure}

Below, in Figs.\ref{g1}-\ref{g3} we depict the first three wave functions $\Psi_{n}^{(q_1,q_2)}$. Notice 
the presence of the deformation in the Fibonacci oscillators is evident in the curves for different $(q_1,q_2)$ 
parameters.

\begin{figure}[htb!]
\centerline{
\includegraphics[{angle=90,height=7.0cm,angle=270,width=7.0cm}]{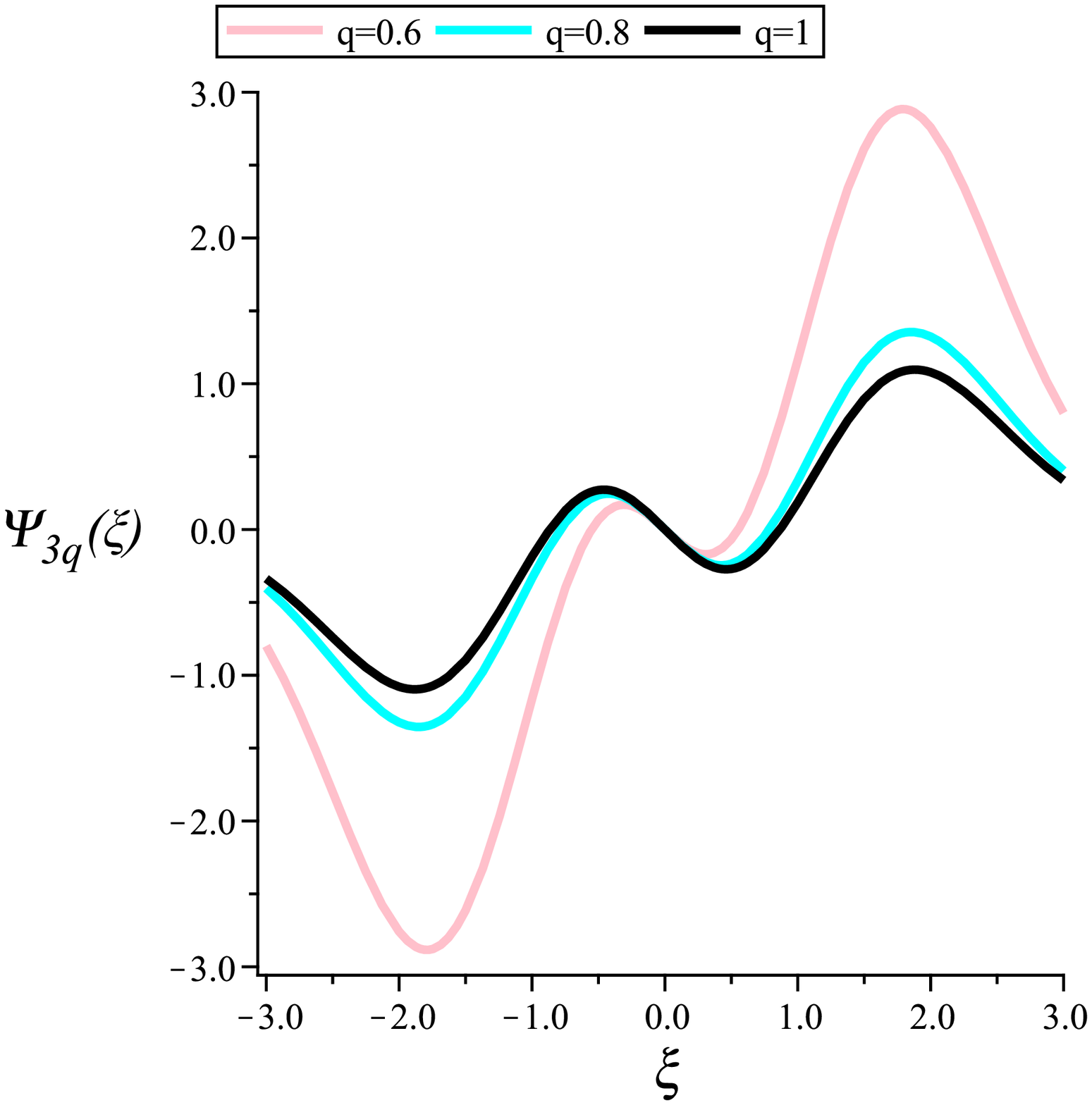}
\includegraphics[{angle=90,height=7.0cm,angle=270,width=7.0cm}]{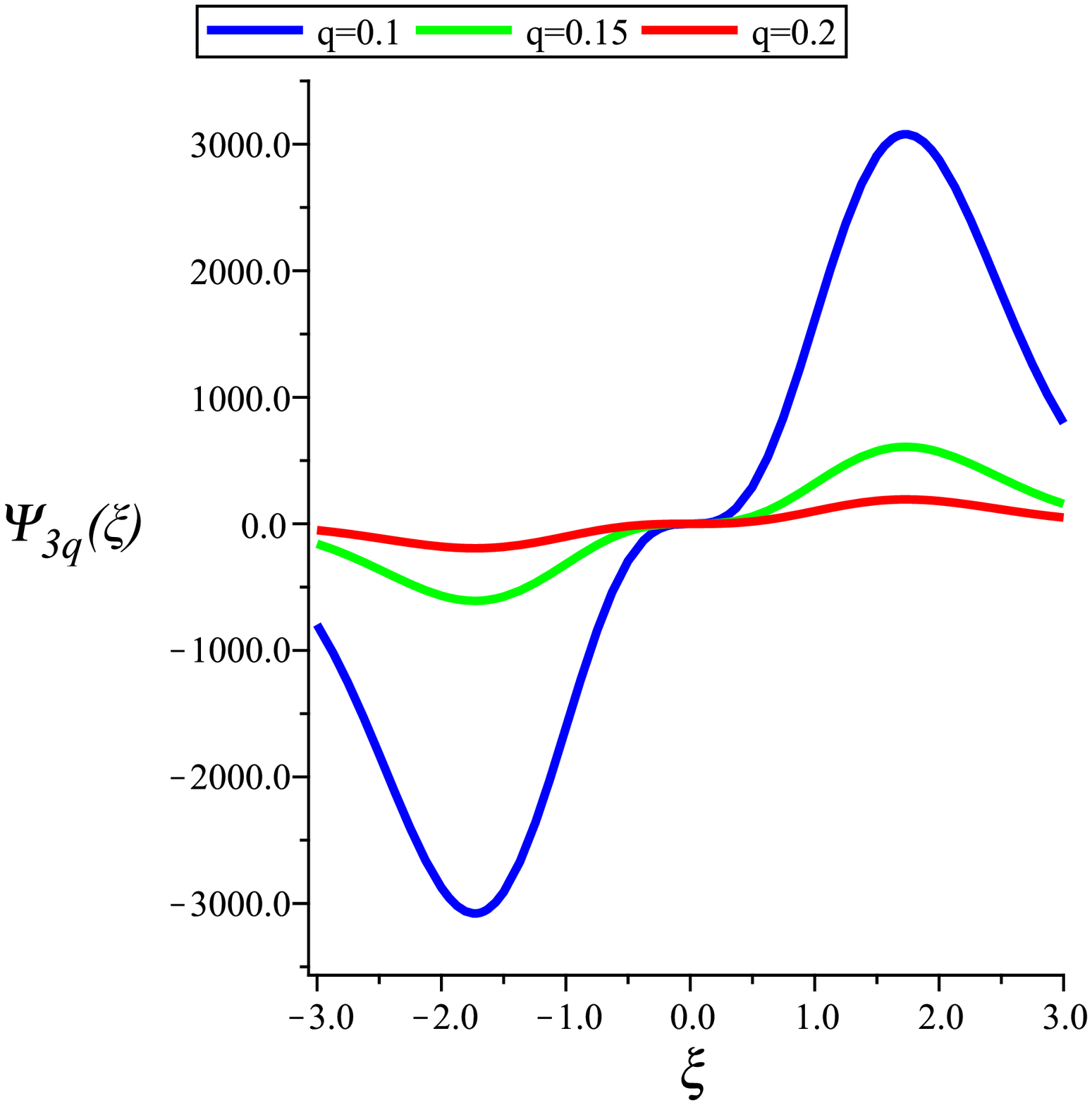}
}\caption{\small{Stationary state $\Psi_{3q}$ for several $(q)$ parameter. The undeformed case is $q=1$ 
(black curve) and the most deformed case is $q_=0.1$ (blue curve).}}
\label{g4}
\end{figure}

We can observe in the Figs.~\ref{g1}-\ref{g3} that the behavior of the curves is altered with the presence 
of the ($q_1, q_2$)-deformation. In all cases the black curves do not present deformation, since 
they are in the limit $q_1 = 1$ and $q_2\to 1$ (or vice versa) and the most deformed case is $q_1 = q_2 = 0.1$ depicted by 
the blue curves. In Fig.~\ref{g1} we can observe that the behaviors are similar. 

The deformation acts simply varying the 
amplitude of the curves and by shifting the positions of nodes of wave functions. 

In Figs.~\ref{g2} and \ref{g3} we note that the presence of the deformation develops a greater role 
in the next excited states. The behavior of the other curves become very different from the black 
curve. On the right panel we have a zoom in three curves to carry it out a better analysis of 
the deformation. {It is worth mentioning that the number of nodes of wave function decreases significantly (by a factor of 2) as the deformation becomes strong enough. This means that in this limit the states are maintained to be even or odd states, but developing less excited modes. As we have anticipated in \cite{bri1}, we can interpret the deformation parameters as impurity (or disorder) factors since the $q$-deformation 
affects the oscillator frequency, which in turn may be associated with the changing of the number of nodes. For further discussion of this phenomenon in deformed diamagnetic material see \cite{bri1} --- and also \cite{gsr} for experimental results.
}

{Damaskinsky and Kulish obtained the $q$-Hermite polynomials \cite{dam}. We can make a little discussion about our results in connection with  $q$-oscillators by reducing  the pair $(q_1,q_2)$ to just $q$ as shown in Eq.~(\ref{eqt1}). Now, we plot in Fig.~\ref{g4} the wave function $\Psi_{3q}$ for some values of $q$.
Notice that  differently from the $(q_1,q_2)$  case as one increases the deformation only the peak amplitudes change. This behavior reinforce the fact that 
$(q_1,q_2)$-deformation develops a more rich physical behavior. 
}
\section{Conclusions}
\label{con}
As expected from Eq.(\ref{eq2.4}), the Fibonacci oscillators have modified behavior in the stationary 
states. It is clear that as the deformation parameters decrease in relation to the 
undeformed case {$q_1 = 1$ and $q_2\to 1$ (or vice versa)}, the {distinct} behavior of the curves becomes more evident. 
This also becomes clear as we look at the Hermite polynomials (\ref{eq2.0}) where $H_3$ feels a 
greater presence of $q_1$ and $q_2$ than $H_1$. We can conclude that the more the states are excited 
the more strong is the deformation on them. This may find interesting applications in quantum 
mechanics such as inclusion of disorders and impurities in the quantum system. For 
instance, in \cite{bri} several studies were put forward uncovering the fact that the $q$-deformation 
affects the oscillator frequency which may be associated with the changing in the strength 
of the `spring constant' associated with such an oscillator as a consequence of introduction 
of impurities or disorders in the system. This should be further addressed elsewhere. 
\section*{Acknowledgments}

We would like to thank CNPq, CAPES, and PNPD/CAPES, for partial financial support.

\end{document}